# Effect of Long-Term Debt on the Financial Growth of Non-Financial Firms Listed at the Nairobi Securities Exchange

David Haritone Shikumo[1], Dr. OluochOluoch[2]&Dr. Joshua Matanda Wepukhulu[3]

*[1-] PhD Finance Candidate*
*Jomo Kenyatta University of Agriculture and Technology, Kenya*
*[2-]Senior Lecturer - Department of Commerce and Economic Studies*
*Jomo Kenyatta University of Agriculture and Technology, Kenya*
*[3-] Lecturer - Department of Commerce and Economic Studies*
*Jomo Kenyatta University of Agriculture and Technology, Kenya*

***Abstract***
*A significant number of the non-financial firms listed at Nairobi Securities Exchange (NSE) have been experiencing declining financial performance which deter investors from investing in such firms. The lenders are also not willing to lend to such firms. As such, the firms struggle to raise funds for their operations. Prudent financing decisions can lead to financial growth of the firm.The purpose of this study is to assessthe effect of Long-term debt on the financial growth of Non-financial firms listed at Nairobi Securities Exchange. Financial firms were excluded because of their specific sector characteristics and stringent regulatory framework.The study is guided by Trade-Off Theory and Theory of Growth of the Firm.Explanatory research design was adopted. The population of the study comprised of 45 non-financial firms listed at the NSE for a period of ten years from 2008 to 2017. The study conducted both descriptive statistics analysis and panel data analysis.The result indicates that Long term debt explains 21.6% and 5.16% of variation in financial growth as measured by growth in earnings per shareand growth in market capitalization respectively. Long term debt positively and significantly influences financial growth measured using both growth in earnings per share and growth in market capitalization.The study recommends that, the management of non-financial firms listed at Nairobi Securities Exchange to employ financing means that can improve the earnings per share, market capitalization and enhance the value of the firm for the benefit of its stakeholders.*

## I. Introduction

Long term debt is part of the financial structure. Financial structure is the way a firm finances its assets through some combination of debt and equity that a firm deems as appropriate to enhance its operations (Kumah, 2013). The determination of a firm's optimal financial structure is vital in deciding how much money should be borrowed and the best mixture of debt and equity to fund business operations (Shubita&Alsawalhah, 2012). Therefore, the choice among the ideal proportion of debt and equity can affect the value of the firm, as well as financial performance.

Long term debt involves strict contractual covenants between the firm and issuers of debt, which is usually associated with high agency and financial distress costs (Tailab, 2014). Equally, Shubita and Alsawalhah (2012) observed that high long-term debt levels in the firm are not conducive for the effective operations of the firm since they increase the risk of bankruptcy. The high debt levels increase the interest rate, which may incapacitate the liquidity levels of the firm (Vermoesen, Deloof&Laveren, 2013). Long term debt is measured as long-term liabilities divided by total assets.

Long term debt is money that is owed to lenders for more than one year from the date of the current balance sheet. Long term debts are the most preferred sources of debt financing among well-established corporate institutions mostly by their asset base, and collateral is a requirement that many of deposits-taking financial institutions (Foyeke, Olusola&Aderemi, 2016). Pelham (2000) argued that long term debts providesmall firms with more competitive advantages when compared with large firms. According to the results, it was found out that there is a direct positive and significant relationship between long term loans and financial performance of small businesses. Pelham (2000) reported that long term debt was positively related to the growth/share/sales effectiveness and gross profit in small and medium-size manufacturing firms. However,

Maniagi, Mwalati, Ondiek, Musiega, and Ruto (2013) noted that long term debt has a weak positive insignificant relationship with ROE. Further, EBaid (2009) revealed that there was no significant relationship between long term debt and financial performance measured using the return on assets.

Financial growth is a measure of efficient utilization of assets by a firm from principal business mode to generate revenues(Aburub, 2012).Financial growth is a general measure of the overall financial health of a firm over a given period(Onaolapo&Kajola, 2010). According to Buvanendra*et al*. (2017), the financial growth of a company is measured as the growth of market capitalization.Market capitalization refers to the total dollar market value of a company's outstanding shares(San &Heng, 2011). Market capitalizationis calculated by multiplying a company's shares outstanding by the current market price of one share(Buvanendra, Sridharan&Thiyagarajan, 2017). The return on investment, return on assets, market value, and accounting profitability reflect financial growth of firms (Ongore&Kusa, 2013).

### 1.1Statement of the Problem

A significant number of the non-financial firms listed at Nairobi Securities Exchange (NSE) have been experiencing declining financial performance which deter investors from investing in such firms (Muchiri, Muturi&Ngumi, 2016). The growth of non-financial firms listed at Nairobi Securities Exchange was 3.7% in 2017 against 4.2% in 2016 (NSE, 2017). Decline in financial performance deter lenders from lending to such firms (Muchiri, Muturi&Ngumi, 2016). For instance, Kenya Airways Limited reported a net loss of Kshs. 26.2bn ($258m) for the financial year 2015-2016, up from Kshs. 25.7bn in the previous financial year(NSE, 2016).Uchumi Supermarket Limited was revived after an agreement between the Kenyan government, suppliers and debenture holders(NSE, 2017).

Muchiri*et al.* (2016)examined the effects of financial structure on financial performance of listed Investment firms in Kenya, and the findings revealed that long term debt had a significant positive relationship with Return on Assets (ROA) and Return on Equity (ROE). The study did not explicitly indicate to what extent, thelong-term debt influencesfinancial performance of listed Investment firms. Further, the study focused only on Investment firms listed at the Nairobi Securities Exchange. The study intends to fill this conceptual gap by focusing on the effect of long-term debt on financial growth of non-financial firms listed at Nairobi Securities Exchange.

### 1.2 Research Objectives

To assess the effect of Long-term debt on the financial growth of Non-financial firms listed at Nairobi Securities Exchange.

### 1.3StatisticalHypothesis

There is no significant effect of Long-term debt on the financial growth of Non-financial firms listed at Nairobi Securities Exchange.

## II. Literature Review

### 2.1 Theoretical Review

The study is guided by Trade-Off Theory and Theory of Growth of the Firm.

#### 2.1.1Trade-Off Theory

Trade-Off Theory postulated by Myers (1984) emphasizes a balance between tax-saving arising from debt, decrease in agent cost and financial distress (Shahar, Shahar, Bahari, Ahmad, Fisal&Rafdi, 2015). Myers (1984) finds that the benefit of the tax shield is offset by the costs of financial distress, and agency cost. In other hard, the optimal level of leverage is achieved by balancing the benefits from interest payments and costs of issuing debt (Jahanzeb, Bajuri, Karami&Ahmadimousaabad, 2014).The balance between tax-saving arising from debt, decrease in agent cost, and financial distress has a significant effect on the financial growth.

Sheikh and Wang (2010) argue that the trade-off theory is expected to choose a target financial structure that maximizes financial growth by minimizing the costs of prevailing market imperfections. The trade-off theory is also referred to as tax-based theories and bankruptcy costs. It assumed each source of money has its own cost and return. These are associated with the firm’s earning capacity and its business as well as insolvency risks (Awan & Amin, 2014). Therefore, a firm with more tax advantage will issue more debt to finance business operations, and the cost of financial distress as well as, benefit from tax-shield are balanced (Chen, 2014).

The purpose of the theory is to explain the fact that firms are mainly financed partly with debt, and partly with equity. It stipulates that there is an advantage to financing with debt, the tax benefits of debt and there is a cost of financing with debt, the costs of financial distress including bankruptcy costs of debt and non-bankruptcy costs. The marginal benefit of further increases in debt declines as debt increases, while the marginal cost increases so that a firm that is optimizing its overall value will concentrate on the trade-off when choosing how much equity and debt to use for financing.

**2.1.2 Theory of Growth of the Firm**

The theory was propagated by Penrose (1959). Penrose argued that firms had no determinant to long run or optimum size, but only a constraint on current period growth rates (Penrose, 1959).According to the theory, financial means for expansion could be found through retained earnings, borrowing, and new issues of stock shares. Retained earnings are one ofthe most important sources to finance new projects in emerging economieswhere capital markets are not well developed. However, firms in the start-up period, when initial investments have not matured yet or whose investment projects are substantially larger than their current earnings, will not have enough financial means from retained earnings and will face a constraint in their growthproject. Firms in this situation may seek external sources of financing; however, the extent of borrowing could be limited by internal factors like high debt-equity ratios that would expose both borrower and lender to increased risk. In other cases, financing of growth projects may be limited by shallow financial markets. Rajan and Zingales (1998) found that industrialsectors with a great need for external finance grow substantially less in countries without well-developed financial markets.

This theory is relevant to this study since it informs the dependent variable which is financial growth. The current studies which have used this theory of firm's growth are; Diaz Hermelo (2007) who conducted a study on the determinants of firm's growth: an empirical examination and Pervan and Višić (2012) who conducted a study on the Influence of firm size on its business success.

**2.2 Conceptual Framework**

Figure 1 shows the conceptual framework.

**Independent Variable** **Dependent Variable**

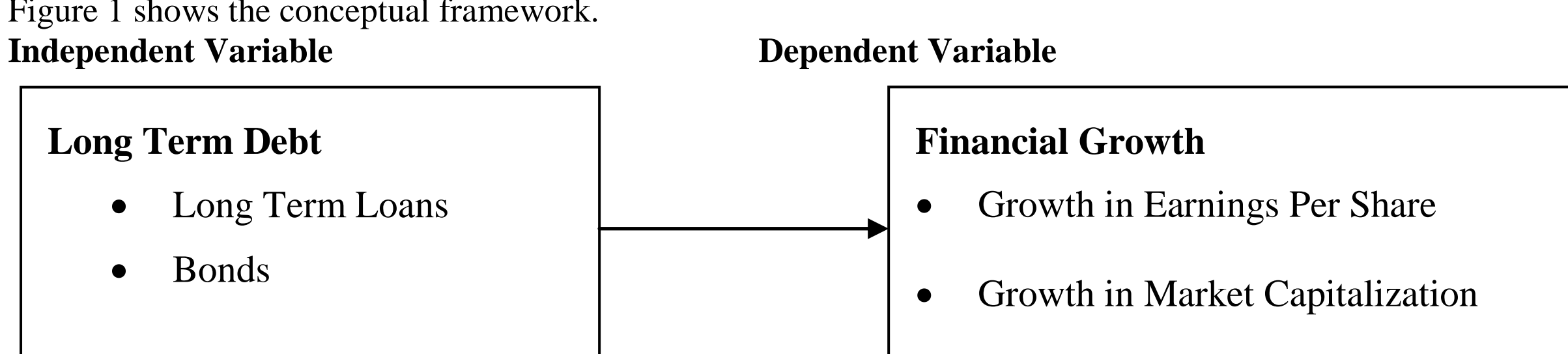

**Figure 1: Conceptual Framework of Long-TermDebt and Financial Growth**

Long-term debt is the independent variable. The dependent variable is the financial growth measured using growth in earnings per share and growth in market capitalization.

## III. Research Methodology

The study adopted positivism research philosophy. Explanatory research design was also adopted. The study conducted both descriptive statistics analysis and panel data analysis. Panel analysis permits the researcher to study the dynamics of change with short time series. The combination of time series with cross-section can enhance the quality and quantity of data in ways that would be impossible using only one of these two dimensions (Gujarati, 2009). The target population of the study comprised 45 non-financial firms listed at the NSE for a period of ten years from 2008 to 31[st]December 2017 (NSE, 2017). Secondary data was extracted from published audited financial statements. Panel data obtained covered a period of 10 years beginning from 2008 and ending in 2017. The panel model to be estimated was: -

$FG_{it}= \beta_0+ \beta_1 LTD_{it}+ \mu$

Where;

FG=Financial growthmeasured by growth in earnings per share and growth in market capitalization of firm i at time t

$\beta_0$ = Alpha coefficient representing the constant term

$\beta_i$ = Beta coefficient

LTD=Long term debt of firm i at time t

i = Firms listed from 2008 to 2017

t = Time period (2008-2017)

$\mu$= Error term

## IV. Research Findings andDiscussions

### 4.1 Descriptive Statistics

Table 1 shows the descriptive statistics forlongtermdebt, earnings per share, and market capitalization.

**Table 1: Descriptive Statistics**

| Variable | Obs | Mean | Std. Dev. | Min | Max |
|---|---|---|---|---|---|
| Long Term Debt | 360 | 0.200195 | 0.186595 | 0.000000 | 1.126967 |
| EarningsPer Share | 360 | 6.468265 | 15.03232 | -46.744 | 100.0483 |
| Market Capitalization in million KES | 360 | 24600.00 | 77300.00 | 116.000 | 721000.00 |

The descriptive results show that the mean value for long-term debt had a mean of 0.200195with a minimum of 0.000000and a maximum of 1.126967. The standard deviation for long term debt was 0.186595. The mean value for earnings per share was 6.468265with a minimum of -46.744, a maximum of 100.0483, and a standard deviation of 15.03232. The mean value for market capitalization as another measure of financial growth wasKshs.24600 million with a minimum of Kshs.116 million and a maximum of Kshs.721,000 million.

### 4.2Trend Analysis

This section presents the analysis of the trends of the variables. The study conducted a trend analysis to establish the movement of the variables overa period of time.Trend analysis for long term debt,earnings per share, and market capitalization are presented in figure 2.

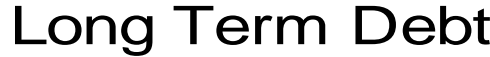

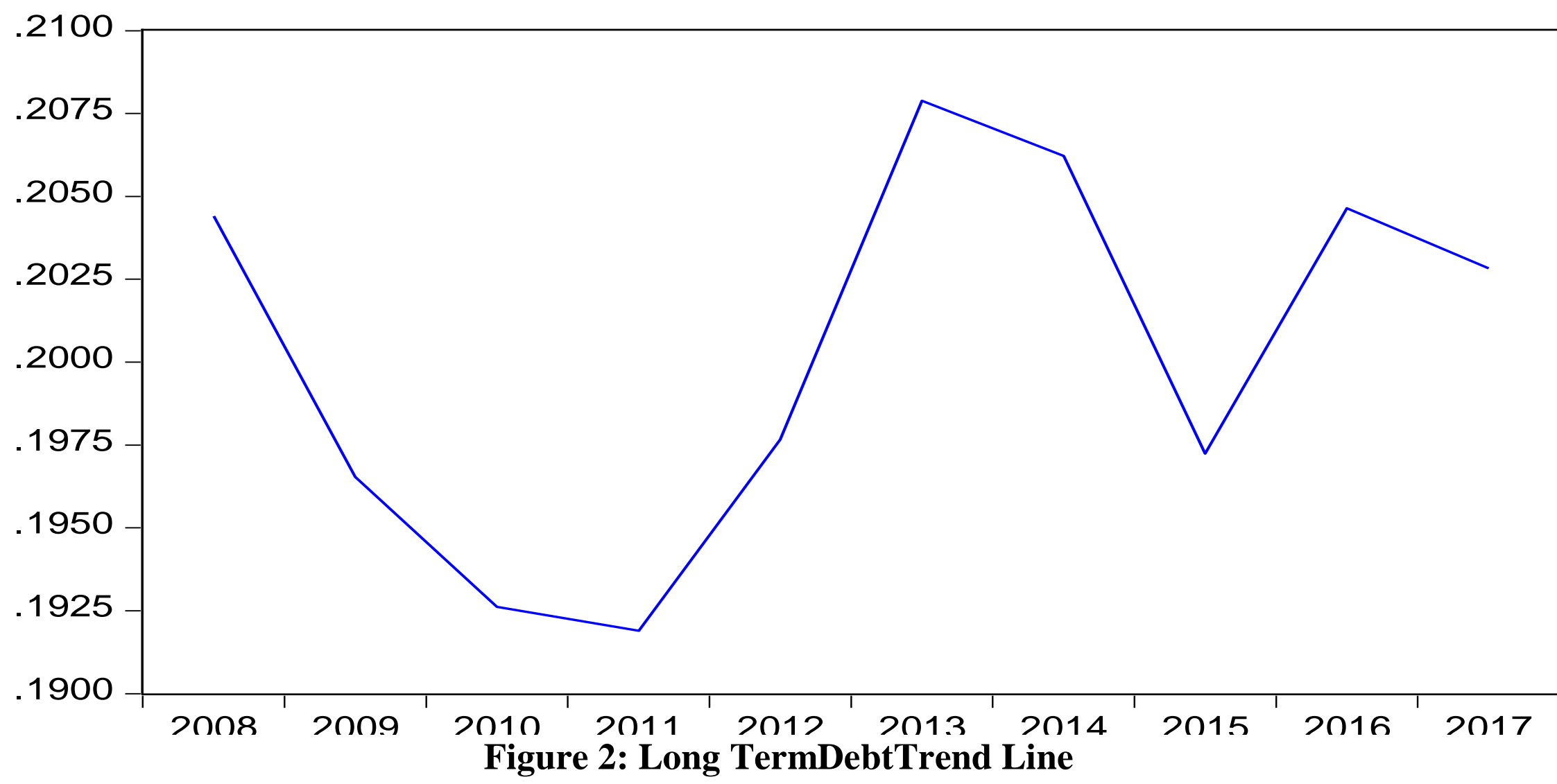

**Figure 2: Long TermDebtTrend Line**

Financing using long term debt dropped from 2008 to the lowest in 2010. However, long term debt rose again from 2011 to the highest in 2013. Long term debt is money that is owed to lenders for more than one year from the date of the current balance sheet. The study by Ebaid (2009) found that there was no significant relationship between long term debt and return on assets. Long term debts are the most preferred sources of debt financing among well-established corporate institutions mostly by their asset base, and collateral is a requirement for many of deposits-taking financial institutions. According to Githire and Muturi (2015), long term debt has a positive and significant effect on the financial performance.

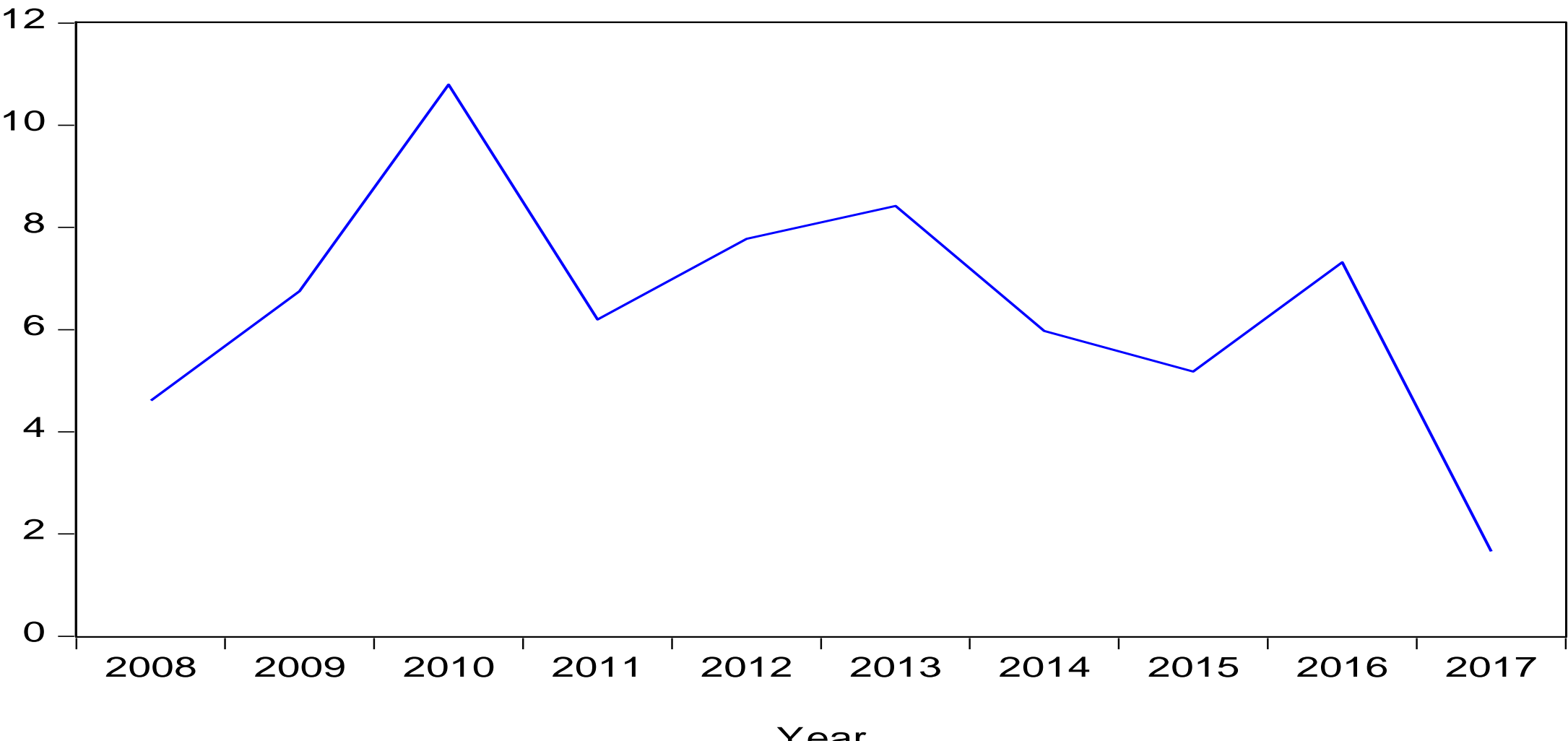

**Figure 3: Earnings Per Share Trend Line**

The trend line shows that earnings per share rose steadily from 2008 to reach the highest in 2010. However, earnings per share dropped further from 2011 to 2012 before rising again in 2013. Earnings per share further dropped drastically to reach lowest in 2017. Earnings per share is considered avital accounting indicator of risk, entity performance, and corporate success. It is used to forecast potential growth in future share price, because changes in EPS are often reflected in share price behavior. Smart and Graham (2012) concur by suggesting that an entity's growth rate is determined by performance indicators such as earnings per share, which is disclosed in the financial statements of companies according to the specifications of the specific accounting standards applied in the respective country. Furthermore, authors have argued that EPS has become a useful investment decision tool for investors because it indicates prospects and growth (Mlonzi, Kruger &Ntoesane, 2011). According to Robbetze, de Villiers, and Harmse (2017), earnings per share correlated best with the changing behavior of share prices.

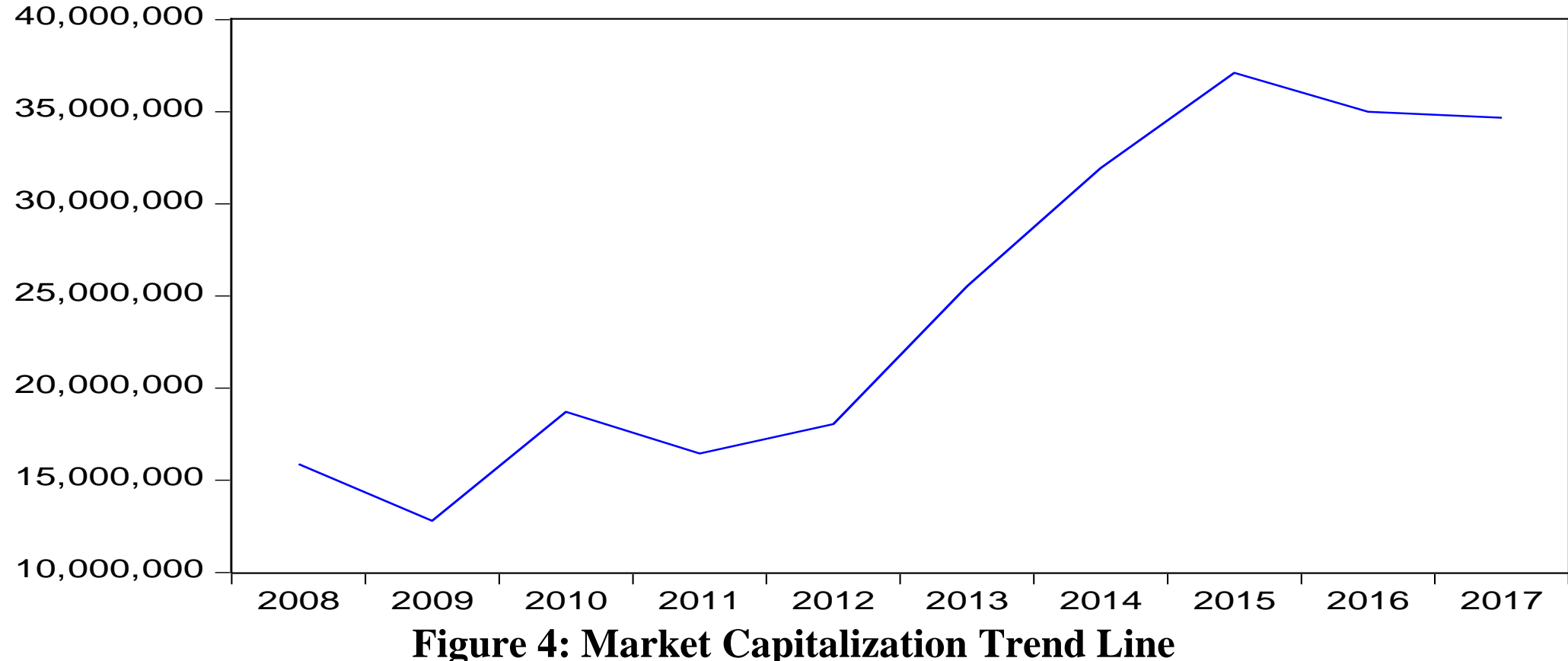

**Figure 4: Market Capitalization Trend Line**

Market capitalization was lowest in 2009 and sharply rose to highest in 2015. Market capitalization is important in projecting the size of an organization because it shows the organization's value. Market capitalization is a measure of the value of companies and stock markets which is an on-going market valuation of a public firm whose shares are publicly traded on a stock exchange computed by multiplying the number of outstanding shares held by the shareholders with the current per share market price at a given time. A market capitalization calculation is a critical part of any stock valuation formula as it represents the total market value of all the company's outstanding shares. Market capitalization can be used as a proxy for the public opinion of a firm's net worth and is a determining factor in some forms of stock valuation. Market capitalization represents the public consensus on the value of a firm's equity. According to Koila, Kiru, and Koima (2018), using the random-effects model also revealed that market capitalization could not be used to predict the outcome of return on equity within the listed firms in the Nairobi Securities Exchange.

**4.3 Correlation Analysis**

The study a conducted correlation analysis between Long term debtand financial growth measured using growth inearnings per share and growth in market capitalization. Table 2 shows the correlation matrix of long-term debtand growth inearnings per share as a measure of financial growth.

**Table 2:Correlation between Long Term Debtand Growth in Earnings PerShare**

| | Growth in EPS | Long Term Debt |
|---|---|---|
| Growth in EPS | 1.000 | |
| Long Term Debt | 0.340 | 1.000 |

The results found that,long term debt and growth in earnings per share are positively associated. Long term debt involves strict contractual covenants between the firm and issuers of debt, which is usually associated with high agency and financial distress costs. High long-term debt levels in the firm's financial structure is not conducive to the effective operation of the firm, since they increase the risk of bankruptcy. According to Opungu (2016), long term debt-equity ratio has a negative effect on return on assets and return on equity but has an insignificant effect on return on capital employed.

**Table 3: Correlation between Long Term Debt and GrowthinMarket Capitalization**

| | Growth in Market Capitalization | Long Term Debt |
|---|---|---|
| Growth in MarketCapitalization | 1.000 | |
| Long Term Debt | 0.333 | 1.000 |

The results found that long term debt and growth in market capitalization arepositively associated. The scarcity of long-term finance is a key impediment to greater investment, and the financial growth. Access to long term financing is one of the critical financial sector policy challenges facing firms. The long-term debt, preferred stock, and common stock together would contribute as the total capital of the firm. The ratio allows the investors to figure out the total risk of investing in a particular business, which can be easily determined by the long-term debt to capitalization ratio. It also shows how financially the company is. The results contrast Lixin and Lin (2008) in a study on the relationship between debt financing and market value of real estate companies of China that long-term borrowing and commercial credit financing has a positive correlation with the firm's market value.

**4.4Effect of Long Term Debton Financial Growth**

Panel regression analysis was conducted on growth in earnings per share and growth in market capitalization.The random model was estimated to determine whether there was a significant relationship between long term debt and variation in the growth in earnings per share. Table 4 presents the panel regression model on long term debt versus growth inearnings per share as a measure of financial growth.

**Table 4: Effect of Long-TermDebt onGrowth in Earning Per Share (EPS)**

| Growth in EPS | Coef. | Std. Err. | T | P>\|t\| | [95% Conf. | Interval] |
|---|---|---|---|---|---|---|
| Long Term Debt | 0.500407 | 0.110795 | 4.52 | 0.000 | 0.283253 | 0.717561 |
| _cons | -1.08621 | 0.926236 | -1.17 | 0.241 | -2.9016 | 0.729182 |
| R-squared: | =0.216 | | | | | |
| Wald chi2(1) | = 20.40 | | | | | |
| Prob> chi2 | =0.0000 | | | | | |

The fitted model from the result is

Growth in EPS = -1.08621+ 0.500407LTD

Where: EPS = Earnings Per Share

LTD = Long Term Debt

As presented in Table4, the coefficient of determination R- Square is 0.216. The model indicates that long-term debt explains 21.6% of variation in the growth inearnings per share as a measure of financial growth. This means 21.6% of the variation in the growth inearnings per share as a measure of financial growth is influenced by long-term debt. The findings further confirm that, the relationship between long term debt and growth in earnings per share as a measure of financial growth is positive and significant with a coefficient of (β =0.500407, p=0.000). It implies that, there exists a positive and significant relationship between long term debt and growth in earnings per share as a measure of financial growth since the coefficient value was positive,and the p-value was0.000< 0.05. It means that, a unitary increase in long term debt leads to growth in earnings per share as a measure of financial growthby 0.500407units holding other elements of financial structure constant.
Long term debts are the most preferred sources of debt financing among well-established corporate institutions mostly by their asset base, and collateral is a requirement for many of the deposits-taking financial institutions. The results do not agree with Salim and Yadav (2012) that long term debt has a negative relationship with earnings per share. Further, Tifow and Sayilir (2015) noted that long-term debt has a significant but negative relationship earnings per share. Fatoki and Olweny (2017) describes earnings per share, as one of the most important financial statistics that is noteworthy for investors and financial analysts. It shows the earnings that a firm has achieved in a fiscal period for an ordinary share and often is used to evaluate the profitability and risk associated with earning and judgments about stock prices. Table 5 presents the regression model on long term debt versus growth of market capitalization as a measure of financial growth.

**Table 5: Effect of Long-TermDebt onGrowth in Market Capitalization**

| Growth in Market Capitalization | Coef. | Std. Err. | T | P>\|t\| | [95% Conf. | Interval] |
|---|---|---|---|---|---|---|
| LongTermDebt | 0.498664 | 0.113902 | 4.38 | 0.000 | 0.27542 | 0.721909 |
| _cons | 2.583273 | 1.284767 | 2.01 | 0.044 | 0.065175 | 5.101371 |
| R-squared: | =0.0516 | | | | | |
| Wald chi2(1) | = 19.17 | | | | | |
| Prob> chi2 | =0.000 | | | | | |

The fitted model from the result is
Growth inMarket Capitalization = 2.583273+0.498664LTD
Where: LTD = Long Term Debt

As presented in Table 5, the coefficient of determination R-Square is 0.0516. The model indicates that, long term debt explains 5.16% of the variations in the growthin market capitalization as a measure of financial growth. It means 5.16%of the variation in the growthin market capitalization as a measure of financial growthis influenced by long term debt. The findings further confirm that, the regression model for long term debtand growthin market capitalization as a measure of financial growthis positive and significant with a coefficient of (β =0.498664, p=0.000). It implies that, there exists a positive and significant relationship between long term debt and growthin market capitalization as a measure of financial growthsince the coefficient value was positive and the p-value 0.000<0.05. It means that a unitary increase in long term debt leads to growth in market capitalization as a measure of financial growth by 0.498664units holding other elements of financial structure constant.The scarcity of long-term finance is a key impediment to greater investment and growth of the firm. Access to appropriate instruments of long-term financing is one of the critical financial sector policy challenges facing firms. The results agree with Lixin and Lin (2008) in a study on the relationship between debt financing and market value of listed real estate companies of China that long-term borrowing and commercial credit financing have a positive correlation with the firm's market value.

**4.5Hypothesis Testing**

The hypothesis was tested using p-value method.The acceptance/rejection criterion was that, if the p-value is greater than the significance level of 0.05, we fail to reject the Ho,but if it's less than 0.05 level of significance, the Hois rejected.The null hypothesis was that,there is no significant effect oflong-term debt on financial growth of Non-financial firms listed at Nairobi Securities Exchange. The results in Table 4 show that, long term debt and growth inearning per share as a measure of financial growthispositively and significantly related withp-value=0.000<0.05. Further, the results in Table 5 shows that, long term debt and growth in market capitalization as a measure of financial growth ispositive and significantly related with p-value=0.000<0.05. The null hypothesis was therefore rejected and concluded that, there is a significant relationship betweenlong-term debt and financial growth of Non-financial firms listed at Nairobi Securities Exchange.

## V. Conclusion

Based on the findings, the study concluded that long term debt has a positive and significant relationship with financial growth measured using growth inearning per share. Long term debts are the most preferred sources of debt financing among well-established corporate institutions mostly by their asset base, and collateral is a requirement for many of the deposits-taking financial institutions. Long term debt involves strict contractual covenants between the firm and issuers of debt, which is usually associated with high agency and financial distress costs. Highlong-term debt levels in the firm are not conducive to the effective operations of the firm since they increase the risk of bankruptcy.

The study also concluded that, long term debt has a positive and significant relationship with financial growth measured using growth inmarket capitalization. Inability of non-financial firms to accesslong-term finance is a key impediment to greater investment and growth of the firm. Access to long-term financing is one of the critical financial sector-wise policy challenges facing firms.

## VI. Recommendations

The study revealed that,long term debt influences financial growth. The study recommends that, the management of non-financial firms listed at Nairobi Securities Exchange to employ financing means that can improve the earnings per share, market capitalization and enhance the value of the firm for the benefit of its stakeholders.